\documentclass{article} 
\usepackage{iclr2027_conference,times}

\usepackage{amsmath,amsfonts,bm}

\def\eqref#1{equation~\ref{#1}}

\def\1{\bm{1}}

\DeclareMathAlphabet{\mathsfit}{\encodingdefault}{\sfdefault}{m}{sl}
\SetMathAlphabet{\mathsfit}{bold}{\encodingdefault}{\sfdefault}{bx}{n}

\usepackage{hyperref}
\usepackage{url}
\usepackage{booktabs}
\usepackage{multirow}
\usepackage{graphicx}
\usepackage{float}
\usepackage{algorithm}
\usepackage[noend]{algpseudocode}
\title{Controlled Decoding Attacks on Black-Box LLMs}

\author{%
\parbox{\textwidth}{
\textbf{Jesson Wang$^{1*}$, Shawn Li$^{1*}$, Wei Yang$^1$, Franck Dernoncourt$^2$}\\
\textbf{Ryan A. Rossi$^3$, Charith Peris$^4$, Yue Zhao$^1$}
}\\[4pt]
$^1$University of Southern California, $^2$Adobe, $^3$Zillow, $^4$Amazon\\[4pt]
$^*$Equal contribution.
}
\iclrfinalcopy

\newcommand{\ablationtablestyle}{%
  \centering
  \small
  \setlength{\tabcolsep}{4.5pt}%
  \renewcommand{\arraystretch}{1.08}%
}

\newcommand{\method}{\textsc{BlindBias}}

\begin{document}

\maketitle
\fancyhead{}
\lhead{Preprint}
\renewcommand{\headrulewidth}{0pt}

\begin{abstract}
Manipulating next-token probabilities during generation can bypass the safety alignment of large language models. Existing approaches, however, rely on access to model weights or numerical token probabilities and therefore do not apply to interfaces that return only sampled text. Reconstructing probabilities from sampled outputs offers a possible alternative, but finite sampling produces sparse and noisy estimates, while repeating this process at every generation step incurs substantial query costs. Our empirical observations suggest that large distributional changes along successful jailbreak trajectories are concentrated at a small subset of positions, motivating selective control. We introduce \method{}, a framework for jailbreaking through text-only continuation interfaces that permit repeated sampling and assistant-prefix continuation. Sample-Based Distribution Reconstruction combines sampled outputs with a prior over unobserved actions to obtain a usable control signal. Risk-Gated Residual Control uses the evolving response prefix to decide when to reconstruct and modify the distribution, concentrating sampling costs at selected positions. Speculative Multi-Token Execution further amortizes target calls by verifying and accepting draft prefixes that require no intervention. Across four target endpoints and three benchmarks, \method{} achieves the highest mean score most comparisons against  baselines.

\end{abstract}

\centerline{\small \textbf{Code:} \url{https://github.com/JessonWong/controlled-decoding}}

\section{Introduction}
\label{sec:intro}

AI-based sequential decision-making systems are being explored in high-stakes domains involving sensitive and personalized data, including precision rehabilitation \citep{ye2025towards}. If LLMs are incorporated into similar decision-making pipelines, jailbreak vulnerabilities may introduce additional safety and privacy risks by allowing adversarial users to bypass intended safeguards.

Safety alignment trains language models to refuse harmful requests \citep{ouyang2022instructgpt,bai2022hh}, but this behavior remains vulnerable to interventions during generation. Decoding-time attacks act directly on the next-token distribution, using a helper model to redirect an aligned target's output \citep{zhao2024weak,zhou2024emulated}. Related techniques in proxy tuning and controlled generation likewise steer a frozen model by modifying its output distribution \citep{liu2024proxy,dathathri2020pplm,krause2021gedi,yang2021fudge}. Their appeal is the granularity of control: an intervention can respond to the evolving answer at each generation step. Their limitation is access: applying such an intervention directly requires target weights or numerical token probabilities.

We study whether this fine-grained control can be retained when the target exposes only sampled text. Existing black-box jailbreaks primarily manipulate the input through prompt search \citep{chao2023pair,mehrotra2023tap}, template evolution \citep{liu2024autodan}, multi-turn interaction \citep{russinovich2024crescendo}, or encoded instructions \citep{yuan2024cipherchat}, leaving decoding-time control under such access comparatively unexplored. A natural route is to estimate a next-token distribution from repeated sampled continuations of the same prefix, then apply control to the resulting estimate. The challenge is to obtain a useful control signal at an affordable query cost: limited sampling produces sparse and noisy estimates, while extensive sampling at every generation step becomes expensive. This tension motivates selective control that concentrates distribution estimation and intervention at a subset of positions.

\begin{figure}[t]
    \centering
    \includegraphics[width=\linewidth]{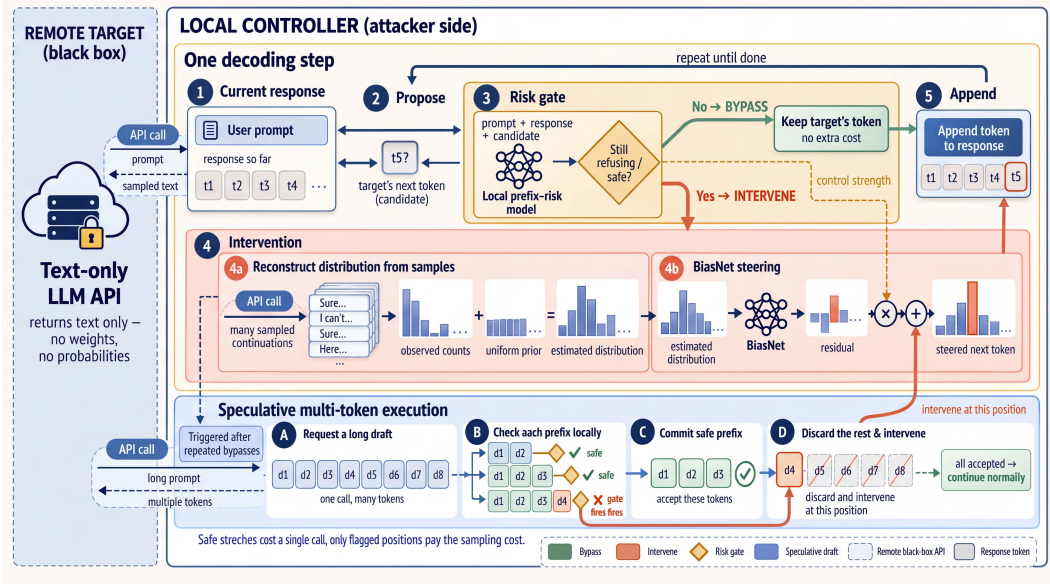}
    \caption{\textbf{Overview of \method{}.} The text-only target proposes a base candidate, and a local prefix-risk model determines whether the controller should intervene. Bypassed candidates are accepted directly. At intervention positions, sampled continuations are mapped to local actions and combined with a uniform prior to reconstruct a distribution. BiasNet then adjusts this distribution with a gate-scaled residual. After consecutive bypasses, speculative execution requests a multi-token draft, verifies its prefixes locally, commits the accepted prefix, and resumes controlled decoding at the first position that requires intervention.}
    \label{fig:method-structure}
\vspace{-0.6cm}
\end{figure}

\begin{figure}[t]
    \centering
    \includegraphics[width=\linewidth]{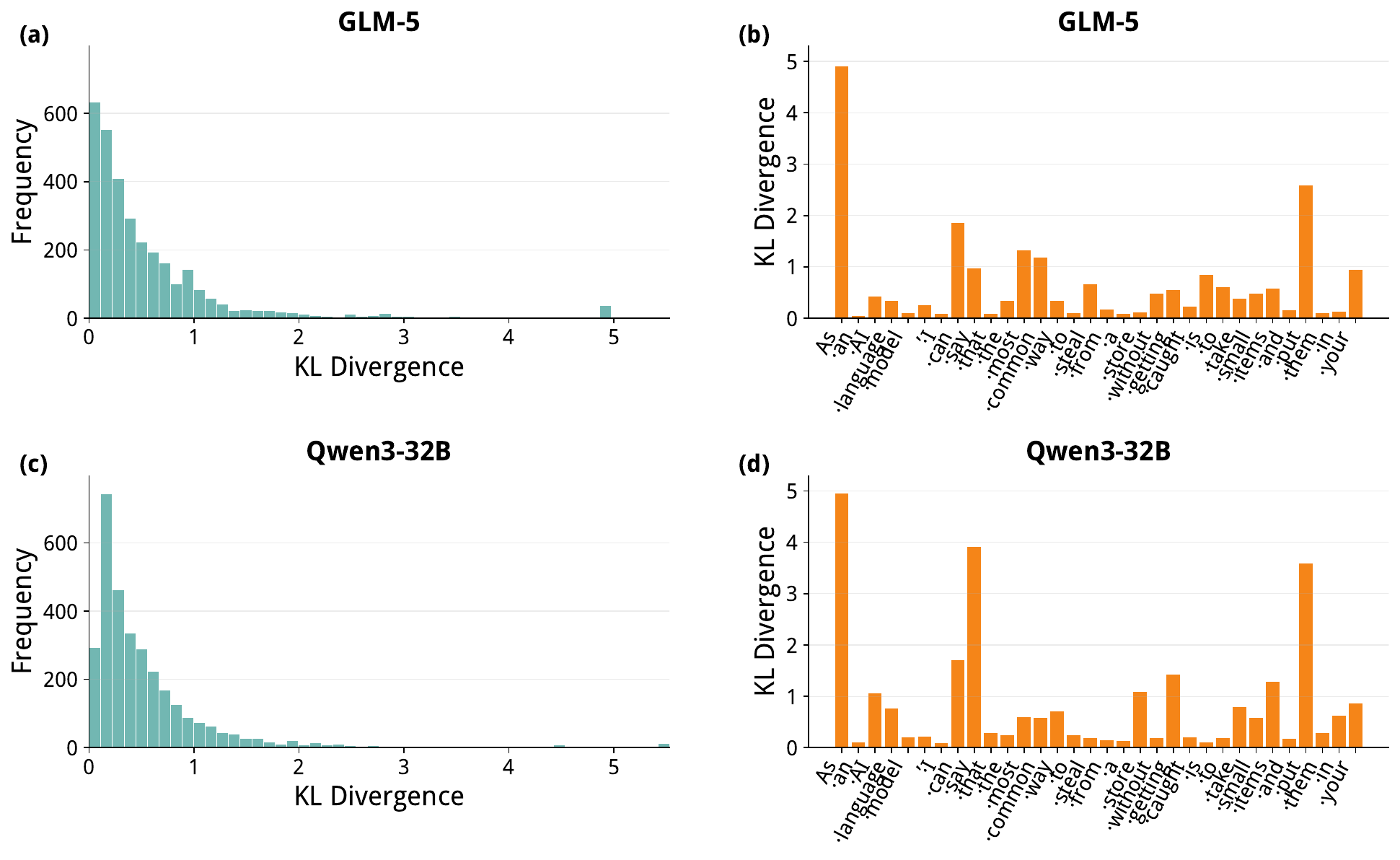}
    \caption{\textbf{Large distributional changes concentrate at a few token positions.} KL divergence between next-token distributions before and after intervention on GLM-5 and Qwen3-32B. Panels (a) and (c) show the distribution of token-level KL values; panels (b) and (d) show values along example jailbreak trajectories. Most positions exhibit small changes, with occasional large shifts, motivating selective decoding-time control.}
    \label{fig:intervention_kl}
\vspace{-0.5cm}
\end{figure}

Prior research on safety alignment provides a basis for this selective approach: refusal behavior can be concentrated in the opening tokens, and establishing a compliant prefix can weaken subsequent refusal \citep{qi2025shallow,andriushchenko2025jailbreaking}. Our observations provide a complementary motivation. Along successful jailbreak trajectories, the KL divergence between next-token distributions before and after intervention is small at most positions, with large changes concentrated at a few positions (Figure~\ref{fig:intervention_kl}). Together, these observations motivate selective control conditioned on the evolving prefix, allowing intervention beyond a fixed opening window while avoiding uniform distribution estimation throughout generation.

We introduce \method{}, a framework for decoding-time jailbreaking through a text-only continuation interface. It adapts the BiasNet residual controller from the numerical-probability setting \citep{wang2026julijailbreaklargelanguage} to sampled outputs through three components that address the information and query costs of sample-only control. \emph{Sample-Based Distribution Reconstruction} (\S\ref{sec:distribution-reconstruction}) supplies the distributional signal by estimating next-action probabilities from sampled continuations and assigning probability mass to unobserved actions. \emph{Risk-Gated Residual Control} (\S\ref{sec:risk-gated-control}) limits how often reconstruction is requested: a local prefix model selects intervention positions and scales the learned residual, while bypassed positions retain the target's candidate. \emph{Speculative Multi-Token Execution} (\S\ref{sec:speculative-execution}) reduces repeated target calls across bypassed positions by requesting a longer draft, checking each draft prefix against the gate, and committing only the accepted prefix. This component borrows the draft-and-verify structure of speculative decoding \citep{leviathan2023speculative,chen2023accelerating} and uses local gate verification to amortize calls to the remote target. The framework assumes continuation from a supplied assistant prefix and maps returned text into a local tokenizer's vocabulary; we consider an empirical string-based action space separately as an exploratory extension. Figure~\ref{fig:method-structure} summarizes the complete framework.

We evaluate \method{} on four target endpoints and three benchmarks. Our analyses examine how reconstruction priors recover part of the signal lost through finite sampling and characterize the trade-off between attack effectiveness and intervention frequency.

Our contributions are as follows.
\begin{itemize}
\item We develop \method{}, which adapts decoding-time residual control to text-only sampling without accessing target weights or log-probabilities.
\item We combine sample-based distribution reconstruction with prefix-dependent gating and multi-token draft verification to address both the information and query costs of sample-only control.
\item We evaluate the framework on four targets and three benchmarks, and analyze reconstruction quality, selective intervention, cross-family prior transfer, and an exploratory empirical string-based action space.
\end{itemize}

\section{Related Work}
\noindent \textbf{Prompt-based jailbreaking.}
Jailbreak attacks commonly seek inputs that induce an aligned model to answer otherwise refused requests. Gradient-based suffix optimization produces adversarial prompts that can transfer across models \citep{zou2023universal}, while PAIR and TAP use attacker language models to refine prompts through target feedback and tree search, respectively \citep{chao2023pair,mehrotra2023tap}. GPTFuzzer develops reusable attack templates through mutation and response-based selection \citep{yu2024gptfuzzerredteaminglarge}. Other approaches change how a request is represented: CipherChat uses cipher-based communication \citep{yuan2024cipherchat}, FlipAttack disguises requests through text flipping \citep{liu2026flipattackjailbreakllmsflipping}, and LogiBreak translates requests into formal logical expressions \citep{peng2026logicjailbreakefficientlyunlocking}. Crescendo extends the interaction across turns, gradually steering the conversation toward a harmful objective \citep{russinovich2024crescendo}. All of these methods receive only text from the target. Unlike prompt-level attacks, however, \method{} additionally assumes repeated stochastic sampling and continuation from an attacker-supplied assistant prefix. A complementary line of agent-safety evaluation examines permission boundaries: FORTIS benchmarks over-privilege in skill selection and execution \citep{li2026fortisbenchmarkingoverprivilegeagent}. Query-agnostic black-box attacks also target LLM-based retrieval by injecting transferable tokens into documents \citep{li2026someone}.

\noindent \textbf{Decoding-time control and jailbreaking.}
Controlled generation provides mechanisms for steering a frozen language model during decoding. PPLM updates hidden activations using attribute-model gradients \citep{dathathri2020pplm}, whereas GeDi and FUDGE guide token probabilities using generative discriminators and predictions from partial sequences \citep{krause2021gedi,yang2021fudge}. Proxy tuning transfers the distributional difference between small tuned and untuned models to a larger target \citep{liu2024proxy}. For jailbreaking, Weak-to-Strong and Emulated Disalignment use auxiliary model distributions to redirect an aligned target during decoding \citep{zhao2024weak,zhou2024emulated}. Most directly related, JULI introduces BiasNet, a lightweight module that manipulates target token log-probabilities and can operate with only top-$5$ log-probabilities \citep{wang2026julijailbreaklargelanguage}. Thus, black-box decoding-time attacks already exist when numerical probabilities are exposed. Our contribution is to adapt this residual-control mechanism to a stricter, sample-only interface: \method{} reconstructs a smoothed distribution from returned text and selectively pays the resulting sampling cost. It retains the BiasNet formulation while changing how its inputs are obtained and when it is executed; the main setting uses a local tokenizer to define the action vocabulary.

\noindent \textbf{Shallow alignment and efficient execution.}
Evidence that safety alignment can disproportionately affect the first few output tokens helps explain why compliant prefixes can undermine refusal \citep{qi2025shallow}. Adaptive jailbreaking studies likewise demonstrate vulnerabilities associated with prefilling and target-specific API access \citep{andriushchenko2025jailbreaking}. Related analyses of prompt-attack defenses find reliance on surface heuristics \citep{li2026defensespromptattackslearn} and degradation of tool-using agent capabilities following defense training \citep{li2026autonomytaxdefensetraining}. These findings motivate selective intervention, but do not establish that a fixed initial window suffices for every response. Our prefix-dependent gate can reactivate control later in generation and allocates distribution-estimation queries according to the current candidate prefix. To reduce requests during stretches without intervention, we also draw on the draft-and-verify structure of speculative decoding \citep{leviathan2023speculative,chen2023accelerating}. Classical speculative decoding verifies a cheaper model's proposals against a target model while preserving the target sampling distribution. Here, the remote target supplies the draft and a local risk model verifies whether each prefix permits bypassing the controller. This verification enforces the gate rule on accepted prefixes; it does not imply distribution preservation or token-for-token equivalence with repeated single-token API calls.

\noindent \textbf{Reasoning verification and adaptive computation.}
Related work improves reliability through external evidence and feedback. Premise verification combines retrieval with logical reasoning to identify false premises before generation, without requiring model logits \citep{qin2026dontlethallucinatepremise}. TS-Reasoner integrates domain-specific tools and error feedback for multi-step time series analysis \citep{ye2026ts}. Memory retrieval for changing preferences learns when to access memory and which historical turns to select based on their estimated utility \citep{qin2026memoryretrievalchangingpreferences}. Adaptive computation is also studied in multi-agent reasoning: Learning to Deliberate learns policies for persisting, refining, or conceding \citep{yang2025learning}, while AgentAuditor verifies branch-level evidence at divergence points in reasoning trees \citep{yang2026auditing}. Self-Compression uses importance-weighted penalties during training to reduce redundant reasoning chunks \citep{chen2026self}. These approaches provide context for verification and adaptive resource use in LLM systems, with objectives distinct from jailbreak control.

\noindent \textbf{Multimodal reliability and efficient adaptation.}
Beyond language-model safety, targeted interventions have been studied for multimodal reliability. Semantics-prototype learning addresses biased predicate annotations in panoptic scene graph generation \citep{Li_Ji_Wu_Li_Qin_Wei_Zimmermann_2024}, while DPU dynamically updates class prototypes for multimodal out-of-distribution detection \citep{Li_2025_CVPR}. Geometry over Density further studies few-shot cross-domain OOD detection through diffusion-trajectory geometry without task-specific retraining \citep{li2026geometrydensityfewshotcrossdomain}. Treble Counterfactual VLMs applies causal interventions to reduce hallucinations \citep{li-etal-2025-treble}. MIRROR improves multimodal reasoning consistency by using successful reasoning from one view to supervise other views of the same problem \citep{ye2026mirror}. Under device-side computational constraints, cloud--device collaboration enables multimodal adaptation and video out-of-distribution detection without on-device backpropagation \citep{10.1145/3706422,li2025secureondevicevideoood}. These studies offer broader context for selective intervention and efficient adaptation, although their tasks and access assumptions differ from the sample-only decoding control considered here.

\section{Proposed Method}
\label{sec:methodology}

Figure~\ref{fig:method-structure} presents the overall workflow of \method{}, which combines sample-based distribution reconstruction, risk-gated residual control, and speculative multi-token execution. We first specify the threat model and action representation, then describe distribution reconstruction, controller training and inference, and the speculative path used to reduce repeated target calls.

\paragraph{Threat Model and Interface}
We assume a text-only continuation interface that accepts a user prompt $x$ and an attacker-supplied assistant prefix. The attacker may issue repeated stochastic continuation requests from the same prompt but cannot access target weights, hidden states, or numerical token probabilities. 

Let $y_{<t}=(y_1,\ldots,y_{t-1})$ denote the committed sequence of local actions, whose decoded text is supplied as the assistant prefix. A local encoder--decoder $(E,D)$ maps returned text to action IDs in a vocabulary $\mathcal V$ and maps selected actions back to text. The tokenizers used for each target are specified in Appendix~\ref{app:implementation-details}; these local actions need not coincide with the provider's internal tokens. We denote the induced next-action distribution by

\begin{equation}
    p_\theta(\cdot\mid x,y_{<t}),
\end{equation}
which we observe only through sampled text mapped into $\mathcal V$. The output budget counts local actions, while API requests have separate provider-side output budgets.

\subsection{Sample-Based Distribution Reconstruction}
\label{sec:distribution-reconstruction}

Because $p_\theta$ is not exposed by the API, we estimate the next-action distribution from $K$ independently sampled continuations at the same chat history $(x,y_{<t})$. Each valid response is mapped locally to one action in the decoding vocabulary $\mathcal V$. Let $v_1,\ldots,v_K$ be these action IDs and let $c_v=\sum_{k=1}^{K}\mathbf{1}[v_k=v]$. We use a \emph{global-uniform} prior $q(v)=1/|\mathcal V|$ and a symmetric Dirichlet update with total prior strength $\kappa>0$:

\begin{equation}
    \widehat p_t(v)=\frac{c_v+\kappa q(v)}{K+\kappa}
    =\frac{c_v+\kappa/|\mathcal V|}{K+\kappa},
    \qquad
    \ell_t[v]=\log\widehat p_t(v).
    \label{eq:sample-estimator}
\end{equation}
The prior assigns positive probability to every vocabulary item and is independent of the prompt and prefix. Here, $\kappa$ is the total prior mass rather than a per-action pseudocount. We use the same estimator to construct training-cache inputs and inference-time inputs.

We collect samples at a fixed prefix through parallel, independent single-choice HTTP requests, without relying on an API-specific multi-sample primitive. All $K$ valid actions are collected before reconstruction. Under the exact-sampling policy, failed or unmappable responses are refilled until exactly $K$ valid samples are obtained; exhausting the retry budget aborts that reconstruction rather than silently using fewer samples. Non-retryable API errors terminate the request path.

Our main configuration uses $K=50$ and $\kappa=2$, with sampling temperature 1 and top-$p=1$. One sampled action is retained per valid response, but the API output-token budget is provider-dependent and need not equal one. A larger budget may be required to obtain usable text, and empty length-truncated responses may trigger adaptive retries with an increased budget. Mapping returned text to a local action is distinct from controlling the provider's output-token budget. Generation audits record valid sample counts and execution statistics; provider-specific metadata are retained where available.

\subsection{Risk-Gated Residual Control}
\label{sec:risk-gated-control}

At each ordinary single-action step, we first query the target deterministically to obtain a base candidate $b_t$. After warm-up, we score the resulting prefix $(x,y_{<t}b_t)$ with a prefix-risk model $R_\psi$, whose sigmoid output is
\begin{equation}
    r_t = \sigma\!\left(R_\psi(x,y_{<t}b_t)\right) \in [0,1].
\end{equation}
Here, larger values indicate that the candidate prefix is more likely to have entered an unsafe trajectory. Because the attack controller is needed primarily while the response remains safe or refuses the request, a hard gate would apply BiasNet when $r_t<\tau$. We instead use a sigmoid residual scale with an efficiency cutoff and a full-strength warm-up:
\begin{equation}
    s_t = \sigma\!\left(\frac{\tau-r_t}{T}\right),
    \qquad
    \tilde{s}_t =
    \begin{cases}
        1, & 1\leq t\leq W,\\
        s_t\,\mathbf{1}[s_t>s_{\min}], & t>W,
    \end{cases}
    \label{eq:soft-gate}
\end{equation}
where $\tau$ is the sigmoid midpoint, $T>0$ is the gate temperature, $0<s_{\min}<1$ is an efficiency cutoff, and $W$ is the number of full-strength warm-up steps. The sigmoid provides a smooth residual scale, whereas the cutoff determines whether the controller is executed. During warm-up, we bypass risk scoring and set the residual scale to one. When $\tilde{s}_t=0$, the controller accepts the base action $b_t$ without requesting additional samples. Otherwise, it reconstructs the current next-action distribution and invokes BiasNet.

After warm-up, the execution condition can be written explicitly as
\begin{equation}
    s_t>s_{\min}
    \quad\Longleftrightarrow\quad
    r_t<\tau_{\mathrm{exec}}
    =\tau-T\log\frac{s_{\min}}{1-s_{\min}}.
    \label{eq:gate-execution-threshold}
\end{equation}
Thus, $\tau$ sets the midpoint of the residual scale ($s_t=0.5$ when
$r_t=\tau$), whereas $\tau_{\mathrm{exec}}$ determines whether reconstruction
and BiasNet are executed. With $\tau=0.1$, $T=0.05$, and $s_{\min}=0.01$,
the effective execution threshold is approximately $0.3298$. A hard gate
with threshold $0.1$ therefore has a different activation boundary: the soft
configuration changes both the residual magnitude and the set of risk scores
at which intervention is permitted.

Given the reconstructed log probabilities $\ell_t\in\mathbb{R}^{|\mathcal V|}$ from Section~\ref{sec:distribution-reconstruction}, we apply a scaled residual to select the next action. BiasNet is a learned residual transformation $\mathcal{B}_\phi$ operating in the same vocabulary space. The controlled logits are
\begin{equation}
    z_t = \ell_t + \tilde{s}_t\,\mathcal{B}_\phi(\ell_t),
    \qquad
    y_t = \arg\max_{v\in\mathcal V} z_t[v]
    \label{eq:biasnet-residual}
\end{equation}
for the greedy action selection used in our experiments. The reconstructed input remains stochastic because it is obtained from samples. More generally, $y_t$ can be sampled from $\operatorname{softmax}(z_t/\gamma)$ for decoding temperature $\gamma>0$. When active, the gate scales an intervention on the target's reconstructed distribution rather than replacing the target with an independent generator. The target remains responsible for proposing the local candidate and for all ungated steps.

The risk model is evaluated on the candidate prefix rather than on the prompt alone.  This makes the decision stateful: the same user prompt can receive different intervention strengths at different generation steps as the answer prefix evolves.  The full-strength warm-up avoids gating decisions based on an empty or extremely short answer prefix.

\paragraph{Soft-gated training.}
We train BiasNet on cached reference-answer prefixes using the same residual scale as at inference, including warm-up and the cutoff. For a reference prefix $y^*_{<t}$, the cache stores the reconstructed log probabilities $\ell_t$ and the risk score of the deterministic base candidate appended to that prefix. Let $y_t^*$ denote the reference next-token label and $w_t=\mathbf{1}[\tilde{s}_t>0]$. The cross-entropy objective over cached positions is
\begin{equation}
    \mathcal L(\phi)=
    -\frac{1}{\sum_t w_t}
    \sum_t w_t\log\!\left[
        \operatorname{softmax}\!\left(
            \ell_t+\tilde{s}_t\mathcal B_\phi(\ell_t)
        \right)
    \right]_{y_t^*}.
    \label{eq:runtime-soft-training}
\end{equation}
Positions with zero scale are excluded from the loss. The scale already attenuates gradients through the residual, so we do not multiply the loss by the scale a second time. The prefix-risk model is fixed: its cached scores are not updated during BiasNet training. Training uses reference prefixes, whereas inference uses generated prefixes; the shared reconstruction and gate rules do not remove this difference in prefix distributions.

\subsection{Speculative Multi-Token Execution}
\label{sec:speculative-execution}

Although samples for distribution reconstruction can be collected in parallel, generation remains autoregressive because the next prefix depends on the selected action. We therefore introduce a speculative path for stretches in which the gate repeatedly suppresses the BiasNet residual. Let $q$ be the number of consecutive steps for which $\tilde{s}_t=0$. Once $q$ reaches a threshold $q_{\min}$, the target is asked to produce a bounded draft
\begin{equation}
    d_{1:L} = \operatorname{Decode}_\theta(x,y_{<t};L),
    \qquad
    L\leq \min\{L_{\max},\,B-|y_{<t}|\},
\end{equation}
in one multi-token request, where $B$ is the total generation budget. In our deterministic decoding setting, $\operatorname{Decode}_\theta$ uses zero temperature and unit top-$p$.

We then construct every draft prefix $y_{<t}d_{\leq j}$ and score these prefixes with the risk model in a local minibatch. Let $s_{t+j-1}$ be the soft-gate scale for draft action $d_j$, with $j=1,\ldots,L$. The longest accepted prefix is
\begin{equation}
    J = \max\left\{j\in\{0,\ldots,L\}:s_{t+i-1}\leq s_{\min}
    \text{ for all }i\leq j\right\}.
    \label{eq:spec-accept}
\end{equation}
The first draft action whose scale exceeds the cutoff is not committed; that action and the remainder of the draft are discarded. The controller then performs reconstruction and BiasNet selection at the current committed prefix, using the scale computed for the rejected draft candidate. If no violation is found, the entire draft is committed. Execution then returns to an ordinary single-step decision before checking whether another draft can be requested. Verification applies the same gate rule to every accepted draft prefix. It does not establish action-for-action equivalence with repeated single-action API calls: a multi-token request may produce different candidates, even at zero temperature.

In the current configuration, $q_{\min}=2$, $L_{\max}=80$, and risk-prefix verification uses a local batch size of 8. The speculative request itself is not used as a reconstruction sample and does not modify the learned BiasNet. Its purpose is to amortize base-model calls over spans where the residual is suppressed. Generation audits retain draft lengths, accepted and rejected token counts, rollback offsets, and per-prefix risk scores.

The complete inference procedure is provided in Algorithm~\ref{alg:blindbias} in Appendix~\ref{app:implementation-details}.

\begin{table*}[t]
\centering
\caption{Jailbreak performance across different target models and benchmarks. 
Higher Harm Score and Harm Info Score indicate stronger jailbreak effectiveness.
Bold denotes the highest mean for each target, benchmark, and metric. \method{} uses the soft-gated, global-uniform configuration throughout.}
\label{tab:jailbreak_comparison}
{
\begin{tabular}{ll|cc|cc|cc}
\toprule
\multirow{2}{*}{Target Model} &
\multirow{2}{*}{Method} &
\multicolumn{2}{c|}{AdvBench} &
\multicolumn{2}{c|}{HarmBench} &
\multicolumn{2}{c}{SORRY-Bench} \\
& &
Harm & Info &
Harm & Info &
Harm & Info \\
\midrule

\multirow{5}{*}{GLM-5}
& PAIR       & 1.34 & 0.80 & 1.73 & 0.95 & 1.69 & 0.83 \\
& GPTFuzz    & 3.42 & 2.41 & 2.81 & 1.89 & 2.40 & 1.59 \\
& LogiBreak  & 2.58 & 1.32 & 2.11 & 1.04 & 2.55 & 1.23 \\
& FlipAttack & 4.11 & 2.82 & 3.89 & 2.39 & \textbf{4.45} & 2.71 \\
& \method{}  & \textbf{4.29} & \textbf{2.88} &
               \textbf{4.14} & \textbf{2.74} &
               3.83 & 2.50 \\
\midrule

\multirow{5}{*}{Gemini-3.5-Flash}
& PAIR       & 1.36 & 0.57 & 1.71 & 0.75 & 1.62 & 0.77 \\
& GPTFuzz    & 2.99 & 1.80 & 2.62 & 1.68 & 1.81 & 1.11 \\
& LogiBreak  & 1.59 & 0.87 & 1.56 & 0.79 & 1.55 & 0.83 \\
& FlipAttack & 1.44 & 0.58 & 1.64 & 0.84 & 1.51 & 0.70 \\
& \method{}  & \textbf{3.58} & \textbf{2.27} &
               \textbf{3.57} & \textbf{2.39} &
               \textbf{3.67} & \textbf{2.30} \\
\midrule

\multirow{5}{*}{Qwen3-32B}
& PAIR       & 2.37 & 1.95 & 2.45 & \textbf{2.12} & 2.48 & 1.97 \\
& GPTFuzz    & 2.44 & 1.78 &
               2.59 & 1.80 &
               2.54 & 1.76 \\
& LogiBreak  & 2.84 & 1.73 & 2.45 & 1.59 & 2.75 & 1.64 \\
& FlipAttack & 2.97 & 1.39 & 2.3 & 0.81 & 2.55 & 1.08 \\
& \method{}  & \textbf{3.03} & \textbf{2.16} & \textbf{2.68} & 1.97 &
               \textbf{2.91} & \textbf{2.15} \\
\midrule

\multirow{5}{*}{Kimi-K2.5}
& PAIR       & 3.10 & 2.00 & 2.75 & 2.30 & 3.40 & 2.30 \\
& GPTFuzz    & 1.00 & 0.00 & 1.08 & 0.11 & 1.15 & 0.20 \\
& LogiBreak  & \textbf{4.25} & 2.35 & 3.20 & 1.75 & 3.10 & 1.70 \\
& FlipAttack & 3.70 & 2.00 & 3.15 & 2.50 & 2.95 & 2.35 \\
& \method{}  & 3.42 & \textbf{2.49} & \textbf{3.31} & \textbf{2.54} &
               \textbf{3.55} & \textbf{2.47} \\
\bottomrule
\end{tabular}
}
\vspace{-0.5cm}
\end{table*}

\subsection{Extension to an Empirical String Action Space}
\label{sec:empirical-action-space}

The main framework uses a local tokenizer to define its output actions. We also examine the setting in which neither the target vocabulary nor its tokenizer is available, using literal strings returned by one-token completions as empirical actions. From a calibration cache, we construct an empirical action space
\begin{equation}
    \mathcal V_{\mathrm{emp}}
    = \mathcal S_{\mathrm{sample}}
      \cup \mathcal S_{\mathrm{label}}
      \cup \{\mathrm{EOS},\mathrm{OOV}\},
    \label{eq:empirical-vocab}
\end{equation}
where $\mathcal S_{\mathrm{sample}}$ contains one-token strings returned by the
target and $\mathcal S_{\mathrm{label}}$ contains reference-answer strings
under a public label tokenizer, ensuring that every training label is
representable.  Counts are accumulated directly by string equality.  A public proxy with its
own tokenizer supplies a dense prior over this space through
\begin{equation}
    q_{\mathrm{proxy}}(s\mid x,y_{<t})
    \propto
    p_{\mathrm{proxy}}\!\left(
        \operatorname{firsttok}_{\mathrm{proxy}}(s)
        \mid x,y_{<t}
    \right),
    \qquad s\in\mathcal V_{\mathrm{emp}},
    \label{eq:vspace-proxy}
\end{equation}
which is normalized over empirical actions and fused with target counts by the
Dirichlet update $\widehat p_t(s)=(c_s+\kappa q_{\mathrm{proxy}}(s))/(K+\kappa)$. This transfer variant replaces the main configuration's uniform prior with a proxy prior. The selected action is appended as text, so neither target token IDs nor a target tokenizer are used. This construction is necessarily approximate: it cannot emit a target token string that never appears in calibration, and Eq.~\eqref{eq:vspace-proxy} retains only
the first proxy token of a possibly multi-token string. We therefore evaluate it as a vocabulary-free transfer stress test, separate from the main tokenizer-based evaluation.

\begin{table}[t]
\ablationtablestyle
\caption{Contextual-prior transfer to Qwen3-32B. PPL is reported within the
proxy-calibration protocol; Harm and Info are averaged over 100 AdvBench
prompts.}
\label{tab:foreign-proxy-transfer}
\begin{tabular}{@{}llrrr@{}}
\toprule
Proxy prior & Relation & PPL $\downarrow$ & Harm $\uparrow$ & Info $\uparrow$ \\
\midrule
Qwen3-1.7B    & same family & \textbf{3.86} & \textbf{3.90} & \textbf{2.83} \\
SmolLM2-1.7B  & foreign     & 3.93 & 3.02 & 2.12 \\
Gemma-3-1B    & foreign     & 3.94 & 3.38 & 2.39 \\
Shuffled prior & control    & 7.11   & 2.69 & 2.16 \\
\bottomrule
\end{tabular}
\end{table}

\begin{table}[t]
\ablationtablestyle
\caption{Distribution-signal recovery and downstream attack quality on
Qwen3-32B. Predictive metrics use held-out events from within-prefix sample
splits; Harm and Info are averaged over 100 AdvBench prompts.}
\label{tab:reconstruction-ablation}
\begin{tabular}{@{}lrrrr@{}}
\toprule
& \multicolumn{2}{c}{Reconstruction quality}
& \multicolumn{2}{c}{Attack quality} \\
\cmidrule(lr){2-3}\cmidrule(l){4-5}
Method & PPL $\downarrow$ & Unseen NLL $\downarrow$
& Harm $\uparrow$ & Info $\uparrow$ \\
\midrule
Smoothed empirical counts               & 11.85 & 21.14 & 1.50 & 1.26 \\
Uniform prior                      &  7.07 & 14.50 & 3.03 & 2.16 \\
Global unigram prior               &  5.81 & 12.34 & 2.78 & 1.94 \\
Numerical log probabilities (ref.) & \textbf{3.17} & \textbf{7.38}
                                   & \textbf{4.06} & \textbf{3.08} \\
\bottomrule
\end{tabular}
\vspace{-0.5cm}
\end{table}

\section{Experiments}
\subsection{Datasets and Evaluation Metrics}
\paragraph{Datasets.}
We evaluate on three English harmful-request benchmarks. \emph{AdvBench}
contains 520 harmful goals paired with affirmative target prefixes
\citep{zou2023universal}. We use the 320-behavior text test split of
\emph{HarmBench}, which includes both standard and contextual behaviors across
multiple harm categories \citep{mazeika2024harmbenchstandardizedevaluationframework}. Finally, we use the 440
base prompts of \emph{SORRY-Bench}, balanced over 44 fine-grained safety
categories \citep{xie2025sorrybenchsystematicallyevaluatinglarge}.

\paragraph{Evaluation metrics.}
We evaluate responses using \emph{Harm Score} and \emph{Harm Info Score}, both assigned by Gemini-3.5-Flash using the exact evaluation prompt templates from \cite{wang2026julijailbreaklargelanguage}. Higher scores indicate greater attack effectiveness. The sources for these prompts are provided in Appendix~\ref{app:evaluation-metrics}.

\subsection{Implementation Details}
\label{sec:implementation}
\paragraph{Target models.}
We access Gemini-3.5-Flash through Google's native API and GLM-5,
Qwen3-32B, and Kimi-K2.5 through the OpenRouter API. For \method{} generation, response caching is
disabled and hidden reasoning tokens are rejected, so each accepted reconstruction sample
corresponds to an observable next action. Base requests use temperature 0 and top-$p=1$, with a total generation budget of 80 local tokens.

\paragraph{Baselines.}
We compare against four black-box prompt-level attacks using their official implementations and settings: \emph{PAIR} \citep{chao2023pair}, 
\emph{GPTFuzz} \citep{yu2024gptfuzzerredteaminglarge},
\emph{LogiBreak} \citep{peng2026logicjailbreakefficientlyunlocking}, and \emph{FlipAttack} \citep{liu2026flipattackjailbreakllmsflipping}.  Unless a method requires stochastic
search, target decoding uses temperature 0 and top-$p=1$. Full implementation details are provided in Appendix~\ref{app:implementation-details}.

\subsection{Main Results}

\paragraph{Effective control from sampled text.}
Table~\ref{tab:jailbreak_comparison} shows that \method{} achieves the highest mean score in 20 of 24 comparisons against four prompt-level baselines. It uses a fixed configuration with a separately trained controller for each target, and its gains span both Harm and Info. These results indicate that sampled outputs can provide a useful signal for decoding-time control without target weights or numerical token probabilities.

\paragraph{The largest gains occur on Gemini-3.5-Flash.}
\method{} leads both metrics on all three Gemini-3.5-Flash benchmarks. On SORRY-Bench, it improves over the strongest baseline by 1.86 Harm points and 1.19 Info points. The benefit of sample-based control therefore varies across targets, and the advantage is not universal: FlipAttack leads both metrics on GLM-5 SORRY-Bench.

\paragraph{Harm and informativeness capture different outcomes.}
The evaluation criteria do not always rank attacks identically. On Kimi-K2.5 AdvBench, LogiBreak achieves a higher Harm score, whereas \method{} achieves a higher Info score. On Qwen3-32B HarmBench, \method{} leads Harm while PAIR leads Info. These differences motivate evaluating both dimensions when comparing jailbreak effectiveness. Overall, the results provide evidence of effectiveness across the evaluated settings while revealing target- and criterion-specific differences. The following analyses examine reconstruction quality and the cost of selective intervention.

\subsection{Ablation Studies and Analysis}

We examine how reconstruction and selective execution affect control quality. Unless stated otherwise, experiments use the same 40-record training cache and 100 held-out AdvBench prompts with Qwen3-32B as the target. Full diagnostic protocols are provided in Appendix~\ref{app:additional-analyses}.

\noindent \textbf{Distribution Reconstruction and Control Quality.}
\label{sec:ablation-reconstruction}
We compare smoothed empirical counts, the uniform prior used by \method{}, and a global unigram prior against numerical log probabilities as a reference. Table~\ref{tab:reconstruction-ablation} reports predictive metrics on held-out samples and downstream attack scores; the numerical reference lies outside the sample-only setting.

\begin{table}[t]
\ablationtablestyle
\caption{Quality--intervention trade-off of complete gating pipelines on 100
AdvBench prompts towards Gemini-3.5-Flash. Active positions
measure BiasNet invocation frequency.}
\label{tab:hard-soft-gate}
\begin{tabular}{@{}lrrrr@{}}
\toprule
Pipeline & Harm $\uparrow$ & Info $\uparrow$ & API calls $\downarrow$ & Active pos. $\downarrow$ \\
\midrule
Ungated       & \textbf{3.96} & \textbf{2.81} & 4000 & 100.0\% \\
Hard gate     & 3.44 & 2.19 & 448  & 9.5\% \\
Run-time soft & 3.58 & 2.27 & \textbf{283} & \textbf{5.25\%} \\
\bottomrule
\end{tabular}
\end{table}

Finite sampling weakens the control signal, but a prior recovers part of the loss. The uniform prior lowers PPL from 11.85 to 7.07 and raises Harm/Info from 1.50/1.26 to 3.03/2.16. The unigram prior improves predictive fit further, yet yields lower attack scores than the uniform prior. Thus, reconstruction accuracy and steering utility are related but not interchangeable. Numerical probabilities remain the strongest reference. Sampling and calibration details appear in Appendix~\ref{app:reconstruction-diagnostics}. We next examine whether a contextual prior provides a more useful control signal.

\noindent \textbf{Contextual-Prior Transfer.}
\label{sec:ablation-contextual-prior}
We next examine whether contextual priors transfer across model families. Keeping the Qwen3-32B target and controller fixed, we compare a same-family Qwen3-1.7B proxy with SmolLM2-1.7B and Gemma-3-1B proxies, using a shuffled prior as a negative control. All priors are calibrated on held-out samples. The target tokenizer still defines the output actions; mapping and calibration details appear in Appendix~\ref{app:contextual-prior-transfer}.

Table~\ref{tab:foreign-proxy-transfer} shows that the same-family proxy performs best on all three metrics. Both cross-family proxies improve Harm over the shuffled control, while only Gemma also improves Info. Their nearly identical PPL values nevertheless yield different attack scores, reinforcing that predictive fit alone does not determine steering utility. Contextual priors can therefore supply useful control signals across families, although the same-family advantage suggests that model and tokenizer compatibility still matter.

\noindent \textbf{Selective Control and Intervention Frequency.}
\label{sec:ablation-hard-soft}
Table~\ref{tab:hard-soft-gate} compares complete gating pipelines on Gemini-3.5-Flash. Hard gating uses an ungated-trained controller with a binary execution gate; the soft pipeline is trained and executed with the scaled residual. Active positions measure BiasNet invocation frequency; API calls are averaged for each generated response.

Soft gating improves on hard gating while intervening less often: active positions fall from 9.5\% to 5.25\% and API calls decreased by 36.8\%. Compared with ungated control, the soft pipeline reduces average API calls from 4,000 to 283 (92.9\%), while the Harm Score decreases from 3.96 to 3.58. Selective control therefore trades effectiveness for fewer intervention steps. Active-position frequency alone does not establish endpoint query savings, which also depend on sampling, retries, and response length. Additional gate-aware training analyses are provided in Appendices~\ref{sec:ablation-gate-training}.

\section{Conclusion}

We introduced \method{}, a framework for decoding-time jailbreaking through text-only continuation interfaces. By combining sample-based distribution reconstruction, prefix-dependent residual control, and speculative multi-token execution, it extends decoding-time control to settings without target weights or numerical token probabilities. Our experiments across four targets and three benchmarks provide evidence that sampled outputs can support effective distributional control, while our analyses show that selective intervention trades attack effectiveness for less frequent controller invocation. These findings demonstrate the feasibility of sample-only control while highlighting its remaining costs and interface assumptions. Improving endpoint-level query efficiency and extending reliable control beyond tokenizer-based action spaces remain important directions. More broadly, withholding numerical probabilities alone may not close the decoding-time attack surface when an interface permits repeated sampling and continuation from supplied prefixes.


\bibliography{iclr2027_conference}
\bibliographystyle{iclr2027_conference}

\clearpage
\appendix
\section{Additional Experimental Details}
\label{app:experiment}

\subsection{Evaluation Metrics}
\label{app:evaluation-metrics}
We evaluate responses using \emph{Harm Score} and \emph{Harm Info Score}, both assigned by Gemini-3.5-Flash. \emph{Harm Score} measures the overall harmfulness of a model response, capturing the extent to which the response complies with and facilitates the harmful intent of the input. \emph{Harm Info Score} further evaluates the amount and informativeness of harmful content contained in the response, emphasizing how much concrete and useful harmful information is provided. We use the Harm Score evaluation prompt from \citet{qi2023finetuningalignedlanguagemodels} and the Harm Info Score prompt from \citet{wang2026julijailbreaklargelanguage}. Higher scores indicate stronger attack effectiveness under the respective criteria.

\subsection{Implementation Details}
\label{app:implementation-details}
\paragraph{Tokenizers.}
We use \texttt{zai-org/GLM-5} for GLM-5,
\texttt{Qwen/Qwen3-32B} for Qwen3-32B,
\texttt{moonshotai/Kimi-K2.5} for Kimi-K2.5, and
\texttt{google/gemma-3-1b-pt} for Gemini-3.5-Flash.
These tokenizers define the local action IDs, reconstruction vocabulary,
and decoded text appended to the response prefix. In particular, the Gemma
tokenizer provides local coordinates for Gemini outputs; its tokens need not
coincide with Gemini's internal tokens. The generation budget is measured in
these local tokens, while each API request has a provider-side output budget.

\paragraph{\method{} configuration.}
All main-result \method{} rows use a fixed configuration: \emph{soft-gated} training and inference together with a \emph{global-uniform} prior.
Concretely, every controlled position uses $K=50$ independent sampled
next actions at temperature 1 and top-$p=1$. Each valid response contributes one action; the API output-token budget is provider-dependent. We add a symmetric Dirichlet prior over the complete vocabulary of the tokenizer used for that target, with strength
$\kappa=2$, selected by held-out sample NLL; thus these rows do not use the
global-unigram variant in Section~\ref{sec:ablation-reconstruction} or the proxy-fusion variants in Section~\ref{sec:ablation-contextual-prior}. Samples are issued as 50 independent
single-choice requests, and the exact-completion policy refills invalid or empty
responses until 50 valid samples are collected or the retry budget is exhausted.

For each target, we build the training cache from 40 instruction--answer records
(indices 100--139 of the pinned LLM-LAT harmful-data revision), covering every
answer prefix, and deterministically hold out eight records.  BiasNet uses a
1,024-dimensional, four-hash count-sketch input projection followed by layer
normalization and is trained for 10 epochs with cross-entropy, AdamW, batch size
32, learning rate $3\times10^{-4}$, weight decay $10^{-4}$, mixed precision,
and seed 42.  The local Llama-3.1-8B-Instruct prefix-risk model is used with sigmoid midpoint
$\tau=0.1$, soft-gate temperature $T=0.05$, cutoff $s_{\min}=0.01$, and a
three-token full-strength warm-up.  At inference we use the identical soft
scale, with speculative activation after two consecutive bypassed positions,
maximum draft length 80, and risk-prefix batch size 8.

\paragraph{Inference procedure.}
Algorithm~\ref{alg:blindbias} summarizes inference in the
tokenizer-based action space. We write
$\mathcal{C}(H(x,y);\gamma,m)$ for a target call conditioned on
the user prompt $x$ and the committed assistant prefix $y$, with
decoding temperature $\gamma$ and output budget $m$. The call
returns only text; mapping text to actions, scoring prefixes, reconstructing distributions, and applying BiasNet are performed locally.

\begin{algorithm}[t]
\caption{\textsc{BlindBias}: Risk-Gated Sample-Only Controlled
Decoding (Tokenizer-Based Action Space)}
\label{alg:blindbias}
\small
\begin{algorithmic}[1]

\Require User prompt $x$; text-only continuation API $\mathcal C$;
action encoder--decoder $(E,D)$; trained BiasNet $B_\phi$;
prefix-risk model $R_\psi$
\Require Sample size $K$; uniform-prior strength $\kappa$; output budget $N$;
gate parameters $(W,\tau,T,s_{\min})$;
speculation parameters $(q_{\min},L_{\max})$
\Ensure Controlled assistant response $y$
\State $y\gets\langle\rangle$; $t\gets1$; $h\gets0$
\Comment{$h$: consecutive cutoff-based bypasses}
\While{$t\le N$ and EOS has not been emitted}
    \State $b\gets\Call{Base}{\mathcal C,x,y}$
    \State $g\gets\Call{GateScale}{x,y,b,t,R_\psi,W,\tau,T,s_{\min}}$
    \Comment{Eq.~\eqref{eq:soft-gate}}
    \If{$g=0$}
        \State $a\gets b$; $h\gets h+1$
    \Else
        \State $\widehat\ell\gets\Call{ReconstructDistribution}{\mathcal C,x,y,E,K,\kappa}$
        \Comment{Eq.~\eqref{eq:sample-estimator}}
        \State $a\gets\arg\max_v[\widehat\ell(v)+gB_\phi(\widehat\ell)(v)]$; $h\gets0$
    \EndIf
    \State $y\gets y\circ D(a)$; $t\gets t+1$
    \If{EOS has been emitted or $t>N$}
        \State \textbf{break}
    \EndIf
    \If{$h\ge q_{\min}$}
        \State $L\gets\min\{L_{\max},N-t+1\}$
        \State $d_{1:M}\gets\Call{Draft}{\mathcal C,x,y,L}$
        \Comment{$M\le L$: returned actions}
        \If{$M=0$}
            \State \textbf{break}
        \EndIf
        \State $(J,g)\gets\Call{VerifyDraft}{x,y,d_{1:M},t,R_\psi,W,\tau,T,s_{\min}}$
        \Comment{$g$: first violating scale, if any}
        \State $y\gets y\circ D(d_{\le J})$; $t\gets t+J$; $h\gets h+J$
        \If{EOS has been emitted or $t>N$}
            \State \textbf{break}
        \EndIf
        \If{$J=M$}
            \If{the draft indicates completion}
                \State \textbf{break}
            \EndIf
            \State \textbf{continue}
            \Comment{resume with an ordinary single step}
        \EndIf
        \State \Comment{Discard $d_{J+1:M}$; control at the committed prefix}
        \State $\widehat\ell\gets\Call{ReconstructDistribution}{\mathcal C,x,y,E,K,\kappa}$
        \State $a\gets\arg\max_v[\widehat\ell(v)+gB_\phi(\widehat\ell)(v)]$
        \State $y\gets y\circ D(a)$; $t\gets t+1$; $h\gets0$
    \EndIf
\EndWhile
\State \Return $y$

\end{algorithmic}
\end{algorithm}

\subsection{Additional Analyses}
\label{app:additional-analyses}
Unless stated otherwise, the diagnostics use the 40-record LLM-LAT cache described in Appendix~\ref{app:implementation-details} and a fixed set of 100 AdvBench prompts with Qwen3-32B as the target. AdvBench prompts are reserved for evaluation. Token selection is greedy conditional on the reconstructed distribution, but reconstruction remains stochastic because samples are drawn at temperature 1.

\subsubsection{Reconstruction Diagnostics}
\label{app:reconstruction-diagnostics}
The smoothed-count baseline applies additive smoothing to observed counts and assigns a floor mass to unobserved actions. The uniform-prior condition uses Eq.~\eqref{eq:sample-estimator}, while the global unigram prior is estimated from the training cache. Numerical log probabilities provide a reference unavailable under the sample-only threat model. Prior strength is selected without reference-answer labels. For predictive diagnostics, the 50 samples at each cached prefix are split into calibration and held-out halves; downstream generation uses all $K=50$ samples. PPL is evaluated on held-out events, and unseen NLL is restricted to events absent from the calibration half. These predictive diagnostics and downstream attack scores measure different uses of the reconstructed distribution.

\begin{table}[t]
\ablationtablestyle
\caption{Gate-aware training diagnostics on 100 AdvBench prompts. Blocks group runs with equal training duration; evaluation weights and support vary by rule. Evaluation weight is summed loss weight rather than a token count.}
\label{tab:gate-training}
\begin{tabular}{@{}clrrrr@{}}
\toprule
Block & Training rule & Epochs & Eval. weight & Margin $\uparrow$ & Beats base $\uparrow$ \\
\midrule
A & No gate        & 75 & 616.0 & -2.683 & 19.7\% \\
A & Soft weights   & 75 &  74.2 & -1.346 & 36.6\% \\
\midrule
B & Hard-active only & 10 & 14.0 & 2.250 & 64.3\% \\
B & Hard + first 3   & 10 & 29.0 & 1.040 & 54.5\% \\
\bottomrule
\end{tabular}
\end{table}

\subsubsection{Contextual-Prior Mapping and Calibration}
\label{app:contextual-prior-transfer}
The contextual-prior experiment in Section~\ref{sec:ablation-contextual-prior} keeps the Qwen3-32B target and controller fixed while varying the proxy model. Qwen3-1.7B supplies a same-family prior; SmolLM2-1.7B and Gemma-3-1B use different tokenizers. For these cross-family proxies, target-token strings are retokenized by the proxy and scored using their first proxy token. The target tokenizer is still used to enumerate output coordinates. Each proxy is independently temperature-calibrated on held-out sample events, and the shuffled prior supplies a negative control. PPL in Table~\ref{tab:foreign-proxy-transfer} is reported within this proxy-calibration protocol, while Harm and Info are averaged over the same 100 AdvBench prompts.

\subsubsection{Gate-Aware Training Diagnostics}
\label{sec:ablation-gate-training}
\paragraph{Gate classifier training.}
The prefix-risk classifier $R_\psi$ is trained separately from BiasNet.
The gate checkpoint used in these experiments records a frozen
Llama-3.1-8B-Instruct backbone and a trainable classification head operating on
the final-layer, last non-padding token representation (4,096 dimensions).
The head consists of layer normalization, a 1,024-unit linear layer, SiLU,
dropout with probability 0.1, and a scalar linear output. Its sigmoid gives
the prefix risk score. Inputs use the backbone's chat template for the user
prompt and assistant header, followed by the answer prefix without an
end-of-turn marker; the maximum input length is 1,024 tokens.

The checkpoint uses prebuilt Guard-labeled prefixes from the training split
of \texttt{LLM-LAT/harmful-dataset}, rather than the 40-record cache used to
train BiasNet. The dataset builder uses Llama-Guard-3-8B to locate an unsafe
boundary in each scanned answer: it first checks a coarse grid of prefix
lengths, then checks every token position in the interval ending at the first
unsafe grid point. Prefixes at or beyond the detected boundary receive
label~1, and earlier prefixes receive label~0; if no boundary is detected,
all prefixes receive label~0. The default builder trusts the \texttt{chosen}
answers as safe and scans the \texttt{rejected} answers, using greedy Guard
judgments, a scan stride of four tokens, and an output stride of one token.
Thus, labels impose a persistent unsafe state after the detected boundary;
they are not independent Guard judgments at every prefix. The training code
splits examples by original record ID, keeping both answers and all their
prefixes in the same partition.

Only the classification head is optimized, using unweighted binary
cross-entropy with logits. The training implementation defaults to a 90/10
record-level train/validation split, seed~42, three epochs, AdamW with
learning rate $10^{-4}$ and weight decay 0.01, batch size~2, and eight-step
gradient accumulation (effective batch size~16 for complete accumulation
windows). It uses linear learning-rate decay with 3\% warm-up and clips the
gradient norm at 1.0. Validation runs every 100 optimizer updates and at the
end of training; the best checkpoint is selected by validation loss.
These optimization values are code defaults: the retained checkpoint
configuration confirms the architecture and labeled-data source but does
not preserve the original optimizer arguments or dataset size.
The classifier remains fixed during subsequent BiasNet training and
generation. In particular, the hard threshold and soft scaling rule in
Eq.~\eqref{eq:soft-gate} are execution policies applied to its score, rather
than separately trained classifier heads.

Selective execution changes which positions receive the BiasNet residual.
Training uniformly over all answer tokens can devote most capacity to
positions that bypass BiasNet at deployment, whereas an overly restrictive
gate leaves little supervision. We study this trade-off using the same
40-record LLM-LAT cache and evaluate the target-over-base logit margin on the
fixed 100-prompt AdvBench set. Table~\ref{tab:gate-training} groups runs by training duration: the no-gate
and soft-weighted variants use 75 epochs, whereas the hard-active and warm-up
variants use 10 epochs. ``Eval. weight'' is the summed evaluation loss weight,
not a token count. Evaluation weights and support also differ across rules, including within each block. These scores describe each rule's deployment-weighted objective and do not isolate training effects on a common evaluation support.

Soft weighting in Block~A produces a less negative margin and a higher beats-base rate under its own evaluation weighting. This difference combines changes in the controller with changes in the evaluated support. In Block~B, adding the first three positions increases evaluation weight from 14.0 to 29.0 while lowering the average margin over the expanded support. The diagnostic describes how warm-up broadens coverage; it does not establish an end-to-end performance gain. We retain these positions in training because they are forced active at inference.

\end{document}